\def\sphinxdocclass{report}
\documentclass[a4paper,10pt,openany,oneside]{sphinxmanual}
\usepackage{iftex}

\ifPDFTeX
  \usepackage[utf8]{inputenc}
\fi
\ifdefined\DeclareUnicodeCharacter
  \DeclareUnicodeCharacter{00A0}{\nobreakspace}
\fi
\usepackage{cmap}
\usepackage[T1]{fontenc}
\usepackage{amsmath,amssymb,amstext}
\usepackage[english]{babel}
\usepackage{times}
\usepackage[Bjarne]{fncychap}
\usepackage{longtable}
\usepackage{sphinx}
\usepackage{multirow}
\usepackage{eqparbox}

\addto\captionsenglish{}
\addto\captionsenglish{}
\SetupFloatingEnvironment{literal-block}{name=Listing }

\addto\extrasenglish{}

\title{Mega Key Authentication Mechanism}
\date{30 October 2015}
\release{7f6735f}
\author{Mega Limited, Auckland, New Zealand \and Guy Kloss \textless{}gk@mega.nz\textgreater{}}
\newcommand{\sphinxlogo}{}
\renewcommand{\releasename}{Release}
\makeindex

\makeatletter
\def\PYG@reset{\let\PYG@it=\relax \let\PYG@bf=\relax%
    \let\PYG@ul=\relax \let\PYG@tc=\relax%
    \let\PYG@bc=\relax \let\PYG@ff=\relax}
\def\PYG@tok#1{\csname PYG@tok@#1\endcsname}
\def\PYG@toks#1+{\ifx\relax#1\empty\else%
    \PYG@tok{#1}\expandafter\PYG@toks\fi}
\def\PYG@do#1{\PYG@bc{\PYG@tc{\PYG@ul{%
    \PYG@it{\PYG@bf{\PYG@ff{#1}}}}}}}
\def\PYG#1#2{\PYG@reset\PYG@toks#1+\relax+\PYG@do{#2}}

\expandafter\def\csname PYG@tok@gu\endcsname{\let\PYG@bf=\textbf\def\PYG@tc##1{\textcolor[rgb]{0.50,0.00,0.50}{##1}}}
\expandafter\def\csname PYG@tok@nv\endcsname{\def\PYG@tc##1{\textcolor[rgb]{0.73,0.38,0.84}{##1}}}
\expandafter\def\csname PYG@tok@k\endcsname{\let\PYG@bf=\textbf\def\PYG@tc##1{\textcolor[rgb]{0.00,0.44,0.13}{##1}}}
\expandafter\def\csname PYG@tok@nl\endcsname{\let\PYG@bf=\textbf\def\PYG@tc##1{\textcolor[rgb]{0.00,0.13,0.44}{##1}}}
\expandafter\def\csname PYG@tok@s1\endcsname{\def\PYG@tc##1{\textcolor[rgb]{0.25,0.44,0.63}{##1}}}
\expandafter\def\csname PYG@tok@vc\endcsname{\def\PYG@tc##1{\textcolor[rgb]{0.73,0.38,0.84}{##1}}}
\expandafter\def\csname PYG@tok@kd\endcsname{\let\PYG@bf=\textbf\def\PYG@tc##1{\textcolor[rgb]{0.00,0.44,0.13}{##1}}}
\expandafter\def\csname PYG@tok@o\endcsname{\def\PYG@tc##1{\textcolor[rgb]{0.40,0.40,0.40}{##1}}}
\expandafter\def\csname PYG@tok@ss\endcsname{\def\PYG@tc##1{\textcolor[rgb]{0.32,0.47,0.09}{##1}}}
\expandafter\def\csname PYG@tok@kr\endcsname{\let\PYG@bf=\textbf\def\PYG@tc##1{\textcolor[rgb]{0.00,0.44,0.13}{##1}}}
\expandafter\def\csname PYG@tok@ni\endcsname{\let\PYG@bf=\textbf\def\PYG@tc##1{\textcolor[rgb]{0.84,0.33,0.22}{##1}}}
\expandafter\def\csname PYG@tok@c\endcsname{\let\PYG@it=\textit\def\PYG@tc##1{\textcolor[rgb]{0.25,0.50,0.56}{##1}}}
\expandafter\def\csname PYG@tok@gt\endcsname{\def\PYG@tc##1{\textcolor[rgb]{0.00,0.27,0.87}{##1}}}
\expandafter\def\csname PYG@tok@si\endcsname{\let\PYG@it=\textit\def\PYG@tc##1{\textcolor[rgb]{0.44,0.63,0.82}{##1}}}
\expandafter\def\csname PYG@tok@ch\endcsname{\let\PYG@it=\textit\def\PYG@tc##1{\textcolor[rgb]{0.25,0.50,0.56}{##1}}}
\expandafter\def\csname PYG@tok@w\endcsname{\def\PYG@tc##1{\textcolor[rgb]{0.73,0.73,0.73}{##1}}}
\expandafter\def\csname PYG@tok@se\endcsname{\let\PYG@bf=\textbf\def\PYG@tc##1{\textcolor[rgb]{0.25,0.44,0.63}{##1}}}
\expandafter\def\csname PYG@tok@ow\endcsname{\let\PYG@bf=\textbf\def\PYG@tc##1{\textcolor[rgb]{0.00,0.44,0.13}{##1}}}
\expandafter\def\csname PYG@tok@cs\endcsname{\def\PYG@tc##1{\textcolor[rgb]{0.25,0.50,0.56}{##1}}\def\PYG@bc##1{\setlength{\fboxsep}{0pt}\colorbox[rgb]{1.00,0.94,0.94}{\strut ##1}}}
\expandafter\def\csname PYG@tok@nc\endcsname{\let\PYG@bf=\textbf\def\PYG@tc##1{\textcolor[rgb]{0.05,0.52,0.71}{##1}}}
\expandafter\def\csname PYG@tok@gs\endcsname{\let\PYG@bf=\textbf}
\expandafter\def\csname PYG@tok@vi\endcsname{\def\PYG@tc##1{\textcolor[rgb]{0.73,0.38,0.84}{##1}}}
\expandafter\def\csname PYG@tok@sh\endcsname{\def\PYG@tc##1{\textcolor[rgb]{0.25,0.44,0.63}{##1}}}
\expandafter\def\csname PYG@tok@kc\endcsname{\let\PYG@bf=\textbf\def\PYG@tc##1{\textcolor[rgb]{0.00,0.44,0.13}{##1}}}
\expandafter\def\csname PYG@tok@nf\endcsname{\def\PYG@tc##1{\textcolor[rgb]{0.02,0.16,0.49}{##1}}}
\expandafter\def\csname PYG@tok@err\endcsname{\def\PYG@bc##1{\setlength{\fboxsep}{0pt}\fcolorbox[rgb]{1.00,0.00,0.00}{1,1,1}{\strut ##1}}}
\expandafter\def\csname PYG@tok@bp\endcsname{\def\PYG@tc##1{\textcolor[rgb]{0.00,0.44,0.13}{##1}}}
\expandafter\def\csname PYG@tok@c1\endcsname{\let\PYG@it=\textit\def\PYG@tc##1{\textcolor[rgb]{0.25,0.50,0.56}{##1}}}
\expandafter\def\csname PYG@tok@cp\endcsname{\def\PYG@tc##1{\textcolor[rgb]{0.00,0.44,0.13}{##1}}}
\expandafter\def\csname PYG@tok@ne\endcsname{\def\PYG@tc##1{\textcolor[rgb]{0.00,0.44,0.13}{##1}}}
\expandafter\def\csname PYG@tok@s\endcsname{\def\PYG@tc##1{\textcolor[rgb]{0.25,0.44,0.63}{##1}}}
\expandafter\def\csname PYG@tok@ge\endcsname{\let\PYG@it=\textit}
\expandafter\def\csname PYG@tok@nb\endcsname{\def\PYG@tc##1{\textcolor[rgb]{0.00,0.44,0.13}{##1}}}
\expandafter\def\csname PYG@tok@kn\endcsname{\let\PYG@bf=\textbf\def\PYG@tc##1{\textcolor[rgb]{0.00,0.44,0.13}{##1}}}
\expandafter\def\csname PYG@tok@gi\endcsname{\def\PYG@tc##1{\textcolor[rgb]{0.00,0.63,0.00}{##1}}}
\expandafter\def\csname PYG@tok@kt\endcsname{\def\PYG@tc##1{\textcolor[rgb]{0.56,0.13,0.00}{##1}}}
\expandafter\def\csname PYG@tok@kp\endcsname{\def\PYG@tc##1{\textcolor[rgb]{0.00,0.44,0.13}{##1}}}
\expandafter\def\csname PYG@tok@gr\endcsname{\def\PYG@tc##1{\textcolor[rgb]{1.00,0.00,0.00}{##1}}}
\expandafter\def\csname PYG@tok@mf\endcsname{\def\PYG@tc##1{\textcolor[rgb]{0.13,0.50,0.31}{##1}}}
\expandafter\def\csname PYG@tok@vg\endcsname{\def\PYG@tc##1{\textcolor[rgb]{0.73,0.38,0.84}{##1}}}
\expandafter\def\csname PYG@tok@mb\endcsname{\def\PYG@tc##1{\textcolor[rgb]{0.13,0.50,0.31}{##1}}}
\expandafter\def\csname PYG@tok@mo\endcsname{\def\PYG@tc##1{\textcolor[rgb]{0.13,0.50,0.31}{##1}}}
\expandafter\def\csname PYG@tok@sc\endcsname{\def\PYG@tc##1{\textcolor[rgb]{0.25,0.44,0.63}{##1}}}
\expandafter\def\csname PYG@tok@nn\endcsname{\let\PYG@bf=\textbf\def\PYG@tc##1{\textcolor[rgb]{0.05,0.52,0.71}{##1}}}
\expandafter\def\csname PYG@tok@mh\endcsname{\def\PYG@tc##1{\textcolor[rgb]{0.13,0.50,0.31}{##1}}}
\expandafter\def\csname PYG@tok@il\endcsname{\def\PYG@tc##1{\textcolor[rgb]{0.13,0.50,0.31}{##1}}}
\expandafter\def\csname PYG@tok@gh\endcsname{\let\PYG@bf=\textbf\def\PYG@tc##1{\textcolor[rgb]{0.00,0.00,0.50}{##1}}}
\expandafter\def\csname PYG@tok@gd\endcsname{\def\PYG@tc##1{\textcolor[rgb]{0.63,0.00,0.00}{##1}}}
\expandafter\def\csname PYG@tok@no\endcsname{\def\PYG@tc##1{\textcolor[rgb]{0.38,0.68,0.84}{##1}}}
\expandafter\def\csname PYG@tok@mi\endcsname{\def\PYG@tc##1{\textcolor[rgb]{0.13,0.50,0.31}{##1}}}
\expandafter\def\csname PYG@tok@sd\endcsname{\let\PYG@it=\textit\def\PYG@tc##1{\textcolor[rgb]{0.25,0.44,0.63}{##1}}}
\expandafter\def\csname PYG@tok@go\endcsname{\def\PYG@tc##1{\textcolor[rgb]{0.20,0.20,0.20}{##1}}}
\expandafter\def\csname PYG@tok@cm\endcsname{\let\PYG@it=\textit\def\PYG@tc##1{\textcolor[rgb]{0.25,0.50,0.56}{##1}}}
\expandafter\def\csname PYG@tok@sb\endcsname{\def\PYG@tc##1{\textcolor[rgb]{0.25,0.44,0.63}{##1}}}
\expandafter\def\csname PYG@tok@nd\endcsname{\let\PYG@bf=\textbf\def\PYG@tc##1{\textcolor[rgb]{0.33,0.33,0.33}{##1}}}
\expandafter\def\csname PYG@tok@na\endcsname{\def\PYG@tc##1{\textcolor[rgb]{0.25,0.44,0.63}{##1}}}
\expandafter\def\csname PYG@tok@sx\endcsname{\def\PYG@tc##1{\textcolor[rgb]{0.78,0.36,0.04}{##1}}}
\expandafter\def\csname PYG@tok@gp\endcsname{\let\PYG@bf=\textbf\def\PYG@tc##1{\textcolor[rgb]{0.78,0.36,0.04}{##1}}}
\expandafter\def\csname PYG@tok@cpf\endcsname{\let\PYG@it=\textit\def\PYG@tc##1{\textcolor[rgb]{0.25,0.50,0.56}{##1}}}
\expandafter\def\csname PYG@tok@s2\endcsname{\def\PYG@tc##1{\textcolor[rgb]{0.25,0.44,0.63}{##1}}}
\expandafter\def\csname PYG@tok@nt\endcsname{\let\PYG@bf=\textbf\def\PYG@tc##1{\textcolor[rgb]{0.02,0.16,0.45}{##1}}}
\expandafter\def\csname PYG@tok@m\endcsname{\def\PYG@tc##1{\textcolor[rgb]{0.13,0.50,0.31}{##1}}}
\expandafter\def\csname PYG@tok@sr\endcsname{\def\PYG@tc##1{\textcolor[rgb]{0.14,0.33,0.53}{##1}}}

\makeatother

\begin{document}

\maketitle
\tableofcontents
\phantomsection\label{index::doc}

\chapter{Mega Key Authentication Mechanism}
\label{preface:mega-key-authentication-mechanism}\label{preface::doc}
For secure communication it is not just sufficient to use strong
cryptography with good and strong keys, but to actually have the
assurance, that the keys in use for it are authentic and from the
contact one is expecting to communicate with.  Without that, it is
possible to be subject to impersonation or man-in-the-middle (MitM)
attacks.

Mega meets this problem by providing a hierarchical authentication
mechanism for contacts and their keys.  To avoid any hassle when using
multiple types of keys and key pairs for different purposes, the whole
authentication mechanism is brought down to a single ``identity key''.

\section{Key Types}
\label{preface:key-types}
A number of key types are used on the Mega platform for different
purposes.
\begin{description}
\item[{Signing/Identity Key (Ed25519).}] \leavevmode
For establishing the identity of a contact and transferring
authenticity to data items via cryptographic signing EdDSA
signatures are used with an Ed25519 elliptic curve signing key
pair \phantomsection\label{preface:id1}{\hyperref[references:ed25519]{\crossref{{[}Ed25519{]}}}} (256 bit key strength).

\item[{Sharing Key (RSA).}] \leavevmode
An RSA key pair (2048 bit key strength) is mainly used for sharing
stored content with contacts.  Additionally, it is used for
establishing a voice/video connection using MEGAchat to avoid MitM
attacks on the WebRTC channel between the peers.

\item[{Chat Key (x25519).}] \leavevmode
For encrypting the sender (session) keys to recipients in MEGAchat,
an elliptic curve Diffie-Hellman (ECDH) approach is used to enable
only one self and the intended recipient access to the key.  For
this an x25519 key pair \phantomsection\label{preface:id2}{\hyperref[references:x25519]{\crossref{{[}x25519{]}}}} (256 bit key strength) is used.

\end{description}

\section{Approaches for Key Authentication}
\label{preface:approaches-for-key-authentication}
A number of approaches are commonly used for ensuring the authenticity
of keys.  The following are used within the Mega platform
\begin{description}
\item[{Key Comparison.}] \leavevmode
Keys are compared between contacts to ensure their authenticity.
Such keys can be lengthy, therefore commonly a cryptographically
secure hash function is used to compute a suitably sized
``fingerprint''.  For comparison of such keys/fingerprints it is
essential to perform an ``out of band'' comparison with the contact,
ideally directly in person with the assurance that no impersonator
is involved.  For people personally not known, other means of
identification (e.g. a photo ID) may be used to assist the process
of establishing the identity.

\item[{Fingerprint Tracking.}] \leavevmode
Once a key has been encountered, its fingerprint is being tracked
(stored privately in tamper-resistant data structure together with
the contact's ID).  Should the case arise that in subsequent
sessions a key's fingerprint is changed, the user can be alarmed of
this issue, and key authentication process can be initiated.

\item[{Key Signing.}] \leavevmode
In case a contact's authenticity of a signing key pair is given, any
item requiring origin authenticity can be verified through a
signature of that item.  In this way public keys can be signed by
the owner's private signing key, and a contact can verify the
signature using the trusted public key.

\end{description}

\section{Key Trust}
\label{preface:key-trust}
The Mega platform does already have some storage provisions for
tracking a user's trust on individual keys. However this concept is
not yet employed or exposed.

\chapter{Key Authentication Tree}
\label{key_authentication:key-authentication-tree}\label{key_authentication::doc}\label{key_authentication:key-authentication}
We are distinguishing between \emph{signed keys} and \emph{unsigned keys.}
Unsigned keys need to be authenticated directly, whereas signed keys
are authenticated indirectly via a signature.  The key used for
signing is an unsigned key authenticated directly.

Authentication of all public keys is arranged in a flat hierarchy.
The identity key is located at the top of the hierarchy (an unsigned
key), which is then used to authenticate all other public keys (signed
keys).

\section{Authentication of Identity Key (Ed25519)}
\label{key_authentication:authentication-of-identity-key-ed25519}
In order to protect oneself against bogus contacts, every key needs to
be authenticated in some fashion.  The main \emph{identity} public key --
an Ed25519 signing key -- is authenticated directly by humans
(unsigned key).  This is commonly done by comparing a large enough
portion of a hash value of the key, which has been obtained via a
cryptographically secure hash function.

A manual fingerprint comparison is to be conducted through a reliable
and trustworthy ``out of band'' channel.  Ideally, this would be through
a personal meeting in real life.  At a minimum this could be done
through a phone conversation with the contact, if one is able to
ensure that the conversation partner on the phone is not an impostor
(e.g. via familiar voice or information items that only that person
may know).  One needs to be aware that the authenticity is only as
good as one is able to prevent any forgeries or impostors.

\section{Authentication of Signed Public Keys}
\label{key_authentication:authentication-of-signed-public-keys}
The identity key is a signing key usable for EdDSA cryptographic
signatures.  Therefore, all further public keys are signed
cryptographically by it, and this resulting signature is made
available to one's contacts.  With this, contacts can verify the
authenticity of any additional public key, once they are sure that the
identity key is authentic.

Even if authenticity of the identity key is not established, one can
at least derive the knowledge that a set of keys is in itself
consistent via verification of the signatures.

The following diagram shows the transitional authenticity
relationships between public keys and the identity key.

\includegraphics{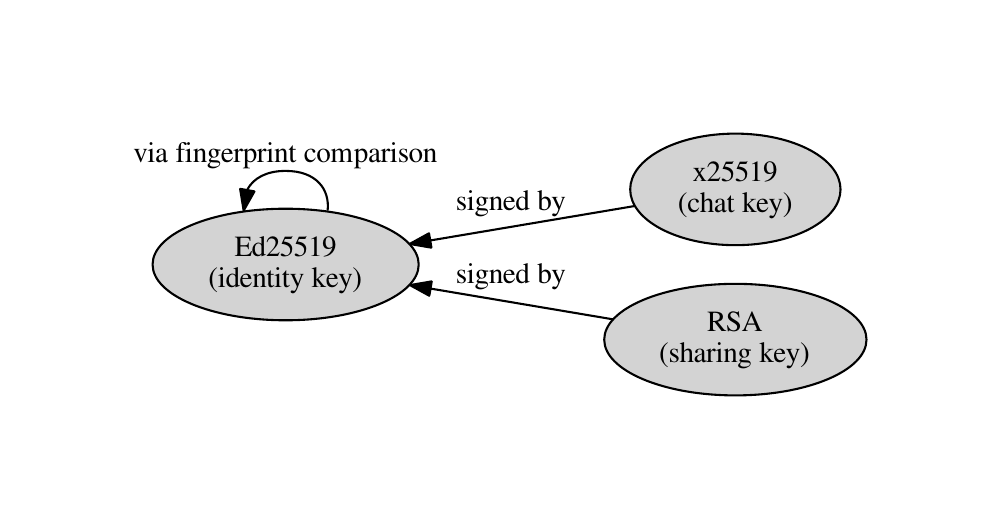}

\section{Tracking of Consistency}
\label{key_authentication:tracking-of-consistency}
Every public key of a key type is tracked in its own \emph{authentication
ring.} The authentication ring contains a key--value list, using a
contact's user handle as a key, and a record of the public key's
fingerprint, authentication method and trust level as a value.

\subsection{``Seen'' Keys}
\label{key_authentication:seen-keys}
If no stronger form of authentication for a key is available, it will
be tracked as ``seen'' for an authentication method.  This allows the
user client to raise an alarm in case the public key unexpectedly
changes, whether it is due to an inconsistency (harmless, but
hindering), or due to an attack attempt on the account of a contact.

\subsection{Fingerprint Comparison of Keys}
\label{key_authentication:fingerprint-comparison-of-keys}
For the identity key we are offering the authentication via
fingerprint comparison.  The Mega client's user interface exposes the
identity key's fingerprints as the ``authenticity credentials'' or
``identity credentials'' within the contact view.  Ed25519 keys approved
with this mechanism will be flagged with the ``fingerprint comparison''
authentication method.  This is the most direct method of key
authentication, but also the most bothersome to users.  Therefore it
is only employed for this particular key.  See
{\hyperref[key_authentication:key\string-fingerprinting]{\crossref{\DUrole{std,std-ref}{Key Fingerprinting}}}} for a description on how these fingerprints
are obtained.

\subsection{Signed Keys}
\label{key_authentication:signed-keys}
The RSA and x25519 public keys are cryptographically signed using the
Ed25519 identity key.  These key signatures are stored as additional
public user attributes.  A contact can retrieve these signature values
along with the public keys and the public Ed25519 key, and verify the
integrity and authenticity of the signed keys.  Such keys are flagged
with the ``signature verified'' authentication method.  This
authentication method causes an almost imperceptible delay for a user
on a particular key and is therefore employed where possible.

\section{Key Fingerprinting}
\label{key_authentication:id1}\label{key_authentication:key-fingerprinting}
A key's fingerprint is a shortened digest computed from a public key,
that is sufficiently long enough as to be taken as a representation of
a particular key that is unfeasible to be forged.  This is commonly
done by computing a large enough portion of a hash value of the key,
which has been obtained via a cryptographically secure hash function.
In Mega's case, this is the slice of the most significant 160 bits of
the SHA-256 hash function, represented in the user interface as 40
hexadecimal characters.  This is the so called ``fingerprint'' of the
public key.

Elliptic curve public keys (x25519 and Ed25519) are hashed directly in
their byte (octet) representation in big-endian format.  For RSA keys,
the byte (octet) representations in big-endian format of the modulus
(\(n = pq\)) and the exponent (\(e\)) are concatenated before
being hashed.

\chapter{Key Authentication Workflow}
\label{workflow::doc}\label{workflow:key-authentication-workflow}
The main goal is to achieve the following for all keys: Obtain all
keys in a way as to not encounter any undetected subversion, and to
achieve maximum tracked authentication for all keys.  As a secondary
goal, we want to achieve the above under the minimum effort.  This
means, that we want to avoid any unnecessary API request/response
round trips as possible.

Obviously caching is a complementary technique that may be employed
for this purpose.  However, it needs to be ensured that cache entries
are not ``stale'' (outdated).  But caching is beyond the scope of this
document and shall not be further discussed here.

As a preliminary, for each key type the corresponding \emph{authentication
ring} needs to be loaded first.  The authentication ring contains key
fingerprints and authentication information for each contact the key
has been loaded previously.  Within the scope of this chapter, it is
assumed that all authentication rings for all keys involved in a
workflow are already loaded, i.e. for loading a chat key the x25519
authentication ring is loaded, as well as possibly the Ed25519
authentication ring when verification of the signature is required.

See Fig. {\hyperref[workflow:fig\string-key\string-load]{\crossref{\DUrole{std,std-ref}{Key loading workflow}}}} for a visual overview of the key loading
procedure outlined below.

\section{Loading of Unsigned Public Keys}
\label{workflow:loading-unsigned-key}\label{workflow:loading-of-unsigned-public-keys}
Loading unsigned public keys (such as the Ed25519 identity key) is
very simple and straight forward.  The contact's public key is loaded,
and its fingerprint is obtained.  If a record for that particular
contact's identity key is already available in the authentication
ring, the fingerprints are compared.  An error dialogue corresponding
to this condition is raised for the case of mismatching fingerprints.
If a key's fingerprint is not tracked, yet, the fingerprint for the
contact is added to a record in the key's authentication ring (with
the authentication method set to ``seen'').

\section{Loading of Signed Keys}
\label{workflow:loading-of-signed-keys}
Loading of signed sub-keys (x25519 and RSA) is more involved,
depending on the availability of signatures (older clients may not
have stored key signatures, yet).

\subsection{Loading Public Keys without Signature}
\label{workflow:loading-public-keys-without-signature}
In the case of the absence of signatures, the key loading mechanism is
similar to the loading of the identity key (see
{\hyperref[workflow:loading\string-unsigned\string-key]{\crossref{\DUrole{std,std-ref}{Loading of Unsigned Public Keys}}}}).  The public key is loaded, and the
fingerprint is compared to that in the corresponding authentication
ring.  In case of mismatches an error dialog is raised, in case of a
missing entry it is added as a tracked ``seen'' key.

\subsection{Loading Public Keys with Signature}
\label{workflow:loading-public-keys-with-signature}
Firstly the public key is loaded, and its fingerprint is obtained
through hashing.  If it is already tracked in the authentication ring,
the common fingerprint comparison is performed.  If this comparison is
failing, the key is not tracked, or it is tracked as only having been
seen, we need to go a step further.  The key's signature as well as
the contact's public identity key are loaded.  Subsequent loading of
the public identity key obviously involves that particular key loading
mechanism (see {\hyperref[workflow:loading\string-unsigned\string-key]{\crossref{\DUrole{std,std-ref}{Loading of Unsigned Public Keys}}}}).  The public key's
signature is verified.  Upon positive validation the public key is
tracked as ``signature verified''.  This may mean that a newly created
signature of a public key may ``upgrade'' the authentication status of a
key from ``seen'' to ``signature verified''.  If the verification fails,
an error dialog is raised to indicate this condition to the user.
\begin{figure}[htbp]
\centering
\capstart

\includegraphics[width=1.000\linewidth]{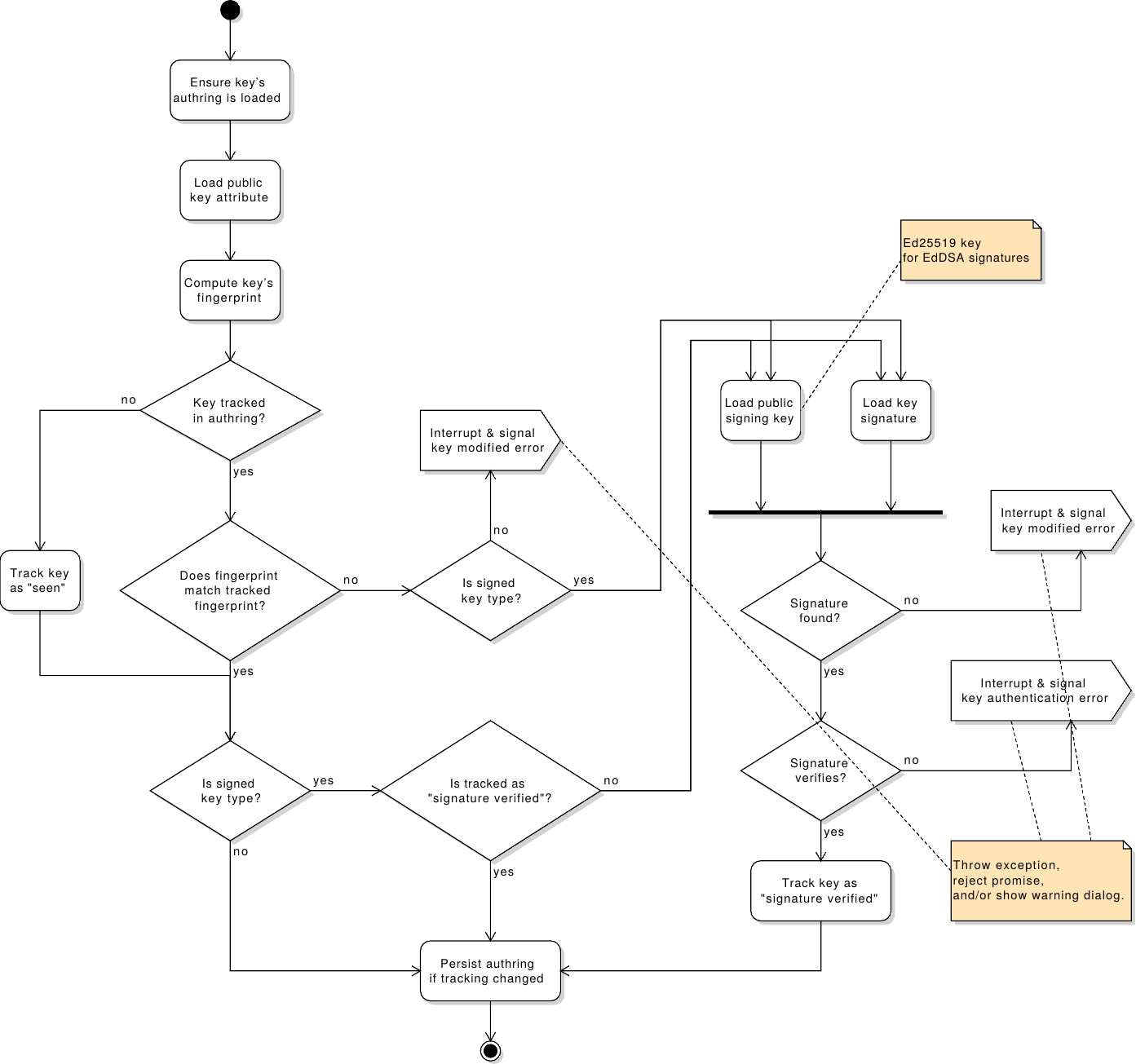}
\caption{Key loading workflow}{\small 
Activity diagram for loading other contacts' public keys along with
key authentication tracking.
}\label{workflow:fig-key-load}\label{workflow:id1}\end{figure}

\section{Own Key Initialisation}
\label{workflow:own-key-initialisation}
When loading one's own keys for operation, the client is to perform
sanity checks, and generate potentially missing keys.  There have been
cases due to bugs and race conditions where some public/private key
pairs were inconsistent with each other, and some clients do not
support all key types, yet.  Therefore a proper procedure for
initialisation and loading is outlined here (see
Fig. {\hyperref[workflow:fig\string-key\string-init]{\crossref{\DUrole{std,std-ref}{Key initialisation workflow}}}}).
\begin{figure}[htbp]
\centering
\capstart

\includegraphics[width=1.000\linewidth]{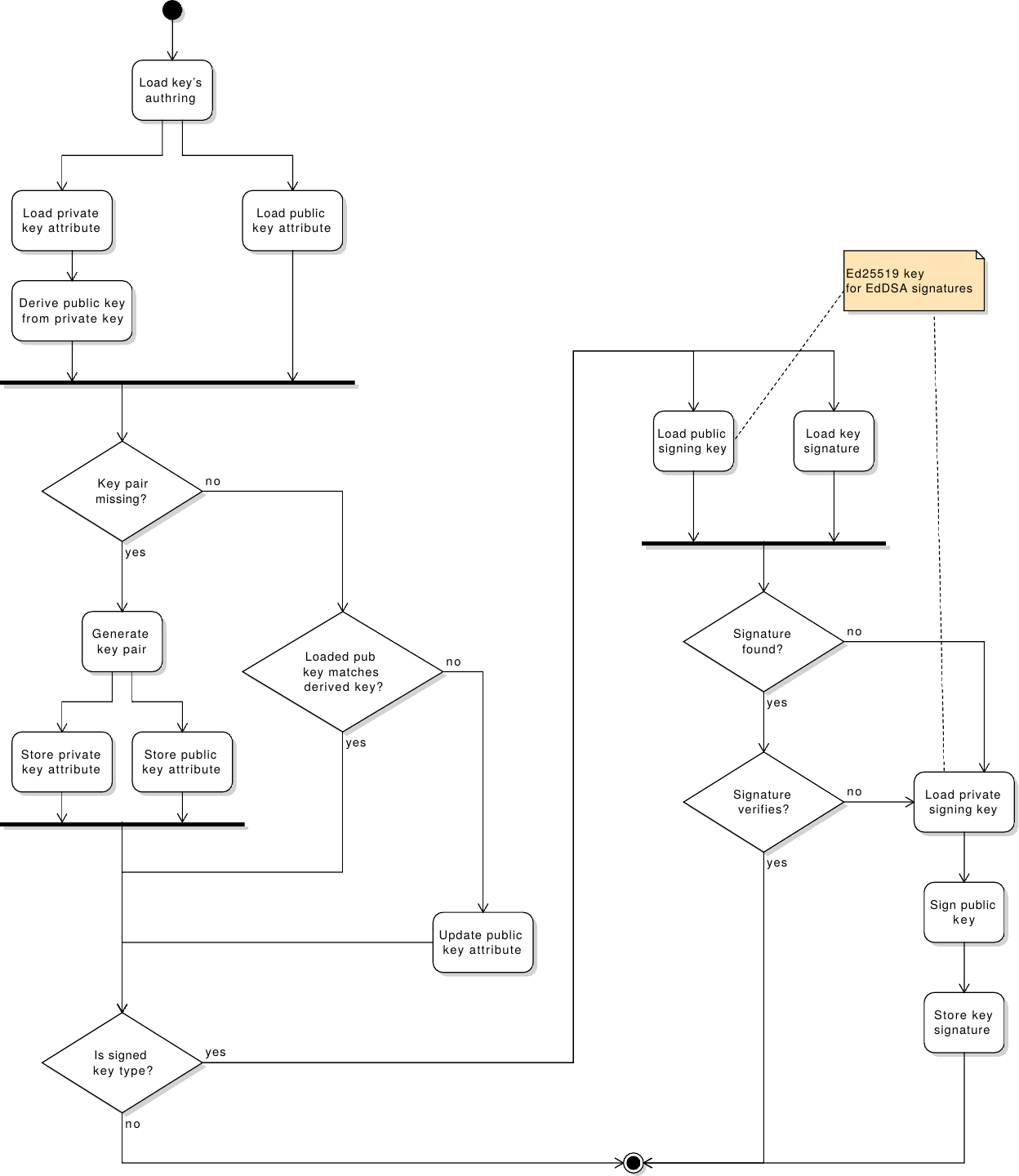}
\caption{Key initialisation workflow}{\small 
Activity diagram for the initialisation and loading of one's own
key pairs, along with resolution of missing attributes or
inconsistencies.
}\label{workflow:fig-key-init}\label{workflow:id2}\end{figure}

\chapter{Future Work}
\label{future_work::doc}\label{future_work:future-work}\begin{description}
\item[{Integration of further keys and key types.}] \leavevmode
Future development may demand other key types beyond the currently
available ones.  This could for example be an OpenPGP key pair.

\item[{Enabling key versioning and rotation.}] \leavevmode
Keys and key pairs may be considered to be ``burnt'' (e.g. over used,
compromised, not considered strong enough) and are in need of being
upgraded by the user.  In these cases old keys need to remain
available (e.g. for old/legacy shares), and keys need to be
explicitly referenced by some version identifier.  Additionally, the
problem of transitive key authentication needs to be solved.

\item[{Key trust.}] \leavevmode
As mentioned initially, we do not expose any mechanisms for trusting
specific keys differently, or for altering such trust levels.  The
key tracking mechanism does already provide a limited attribute for
this, but it is not used, yet.

\item[{Web of Trust.}] \leavevmode
From the current vantage point a \emph{Web of Trust} (WoT) is an
interesting idea.  However, it is mostly not exercised ``in the wild''
(e.g. with OpenPGP).  We are not going into the trouble of
discussing any issues further here, but it seems unlikely that a WoT
approach will be implemented on top of the Mega platform.

\end{description}

\renewcommand{\indexname}{Index}
\printindex
\end{document}